\documentclass[12pt]{article}
\usepackage{amsfonts}
\usepackage{amsmath}
\usepackage{amssymb}
\usepackage{graphicx}
\usepackage{color}
\usepackage[all, knot]{xy}
\usepackage{tikz}
\usepackage{cancel}
\usepackage{comment}

\usepackage{ulem}
\usepackage{hyperref}

\usepackage[utf8]{inputenc}
\usepackage{epstopdf}
\usepackage[footnotesize]{caption}
\usepackage{amsthm}
\usepackage{enumitem}
\usepackage{mathrsfs}

\makeatletter
\newcommand\xleftrightarrow[2][]{%
	\ext@arrow 9999{\longleftrightarrowfill@}{#1}{#2}}
\newcommand\longleftrightarrowfill@{%
	\arrowfill@\leftarrow\relbar\rightarrow}
\makeatother

\usepackage[margin=3cm]{geometry}

\def \be {\begin{equation}}
\def \ee {\end{equation}}
\def \bea {\begin{eqnarray}}
\def \eea {\end{eqnarray}}
\def \nn {\nonumber}

\def \rr {\raise.35ex\hbox{\small $\prime$}\kern-.17em{\mbox{\large $\imath$}}}

\def \dels {\partial\kern-.6em /\kern.1em}
\def \As {{A\kern-.5em / \kern.5em}}
\def \Ds {D\kern-.7em / \kern.5em}

\def \ks {k\kern-.5em /}
\def \ls {l\kern-.5em /}

\newcommand{\ci}[1]{}

\newcommand{\ba}{\begin{eqnarray}}
\newcommand{\ea}{\end{eqnarray}}
\newcommand{\bal}{\begin{align}}
\newcommand{\eal}{\end{align}}
\newcommand{\bay}[1]{\left(\begin{array}{#1}}
\newcommand{\eay}{\end{array}\right)}

\newcommand{\hide}[1]{}

\DeclareMathOperator{\Tr}{Tr}

\newlist{axioms}{enumerate}{2}
\setlist[axioms,1]{label=\textbf{A\arabic{axiomsi}.}, ref=A\arabic{axiomsi}}
\setlist[axioms,2]{label=\textbf{A\arabic{axiomsi}\rlap{\myEnumCounter{axiomsii}}.},%
                   ref=A\arabic{axiomsi}\myEnumCounter{axiomsii},%
                   align=parleft,%
                   leftmargin=0em,%
                   itemsep=1.4ex,%
                   before={\stepcounter{axiomsi}}}

  \usetikzlibrary{decorations.markings}

\begin{document}

\begin{titlepage}
\begin{center}

\textbf{\LARGE
Emergence of Time from\\ 
Unitary Equivalence
\vskip.3cm
}
\vskip .5in
{\large
Pak Hang Chris Lau$^{a,b,c}$ \footnote{e-mail address: phcl2@panda.kobe-u.ac.jp} and
Chen-Te Ma$^{d,e}$ \footnote{e-mail address: yefgst@gmail.com}
\\
\vskip 1mm
}
{\sl
$^a$
Department of Physics, Osaka University, Toyonaka, Osaka 56, Japan.
\\
$^b$
Department of Physics, Kobe University, Kobe-shi 657-8501, Hyogo, Japan. 
\\
$^c$
National Center for Theoretical Sciences, National Tsing-Hua University,\\
Hsinchu 30013, Taiwan, R.O.C.
\\
$^d$
Department of Physics and Astronomy, Iowa State University, Ames, Iowa 50011, US.
\\
$^e$
Asia Pacific Center for Theoretical Physics,\\
Pohang University of Science and Technology, 
Pohang 37673, Gyeongsangbuk-do, South Korea. 
}\\
\vskip 1mm
\vspace{40pt}
\end{center}

\newpage
\begin{abstract} 
We discuss the concept of unitary equivalence $\hat{H}\sim\hat{U}^{\dagger}\hat{H}_{\mathrm{mod}}\hat{U}$  between the modular Hamiltonian $\hat{H}_{\mathrm{mod}}$ and the subsystem Hamiltonian $\hat{H}$ in the context of realizing the emergence of time through a unitary operator $\hat{U}$. 
This concept suggests a duality between the modular flow and time evolution. 
Additionally, requiring unitary equivalence implies a connection between the "Modular Chaos Bound" and the "Chaos Bound". 
Furthermore, we demonstrate this duality using quantum chaos diagnostic quantities in the thermofield double state of a fermionic system. 
Quantum chaos diagnostic quantities are mathematical measures that characterize chaotic behavior in quantum systems. 
By examining these quantities in the thermofield double state, we illustrate the duality between them and the modular Hamiltonian. 
We show a specific duality between correlators, the spectral form factor, and the Loschmidt echo with the modular Hamiltonian. 
The spectral form factor is a quantity that provides information about the energy spectrum of a quantum system, while the Loschmidt echo characterizes the sensitivity of a system's modular time evolution to perturbations. 
Finally, we demonstrate that a different entanglement spectrum does not impose the same constraint on the subsystem Hamiltonian. 
The entanglement spectrum is related to entanglement entropy and provides information about the eigenvalues of the reduced density matrix associated with a subsystem. 
We discuss complex concepts related to the interplay between quantum chaos, time emergence, and the relationship between modular and subsystem Hamiltonians. 
These ideas are part of ongoing research in quantum information theory and related fields. 
\end{abstract}
\end{titlepage}

\section{Introduction}
\label{sec:1}
\noindent
The out-of-time ordered correlator (OTOC) is a quantity that is used to study Quantum Chaos and Quantum Information. 
It measures the sensitivity of a quantum system to small changes in its initial conditions \cite{Liao:2018uxa}. 
In the context of holography, the exponent of the OTOC, denoted as $\lambda$, has been identified as the quantum analog of the classical Lyapunov exponent. Saturation of a bound, $\lambda=2\pi/\beta$, where $\beta$ is the inverse temperature of the boundary theory, in the OTOC can imply the emergence of Einstein gravity in the bulk theory \cite{Maldacena:2015waa}. 
\\

\noindent
The spectral form factor (SFF) is another quantity used to probe Quantum Chaos. 
It provides information about the statistics of energy levels in a quantum system. 
Random matrix theory has been employed to study the SFF in the context of black holes, offering insights into late-time behavior and constraints on the bulk theory \cite{Dyer:2016pou,Cotler:2016fpe}. 
\\

\noindent
The question of reconstructing the bulk geometry from the entanglement spectrum of the boundary theory is still an open area of research. 
This is analogous to the famous question in mathematics, "Can one hear the shape of a drum?". 
In the Anti-de Sitter (AdS)/Conformal Field Theory (CFT) correspondence, it is believed that there exists a one-to-one correspondence between a gravitational theory in the bulk (AdS space) and a quantum field theory on the boundary. However, the holographic principle alone may not provide enough information to uniquely determine the bulk geometry. 
The idea behind our question is that if there exists a unitary transformation, denoted as $\hat{U}_{\cal O}$, which leaves an observable invariant
\bea
\hat{H}\propto \hat{U}_{\cal O}^{\dagger}{\hat{\cal O}}\hat{U}_{\cal O}, \ 
|n\rangle\rightarrow\hat{U}_{\hat{\cal O}}|n\rangle,
\eea
where $|n\rangle$ is an eigenstate of $\hat{H}$, and $\hat{H}$ is the Hamiltonian, then the corresponding bulk geometry should also be insensitive to this unitary transformation and only depend on the spectrum of the Hamiltonian $\hat{H}$. 
This concept is referred to as unitary equivalence. 
\\

\noindent
We propose considering $\hat{{\cal O}}$ as a modular Hamiltonian, given by 
\bea
\hat{H}=\beta\hat{U}^{\dagger}{\hat{\cal O}}\hat{U},    
\eea
where $\beta$ is the inverse temperature and $\hat{H}$ is now the subsystem Hamiltonian. 
The entanglement spectrum of the boundary theory then constrains the form of the Hamiltonian \cite{Jafferis:2020ora}. 
The central question in this letter is: {\it Can one hear the shape of geometry from unitary equivalence?} 
The central question we raise is whether it is possible to extract information about the shape of the bulk geometry solely from the concept of unitary equivalence. 
In other words, can the entanglement spectrum alone provide enough information to reconstruct the bulk geometry in the AdS/CFT correspondence? 
\\

\noindent
The Loschmidt echo with a modular Hamiltonian (LEMH) is a mathematical tool used to analyze the behavior of quantum systems and provides a bound on chaotic behavior. 
The modular chaos bound \cite{DeBoer:2019kdj}, derived from the LEMH, relates to the sensitivity of a quantum state to infinitesimal perturbations. 
When this bound is saturated, it implies that the quantum state is maximally sensitive to such perturbations.
\\

\noindent
The unitary equivalence condition suggests that the chaos bound and the modular chaos bound are equivalent when the modular parameter is related to time by the formula $s= t\beta$. 
The relationship between the modular parameter and time, known as the "Thermal Time Hypothesis" \cite{Connes:1994hv}, suggests that modular evolution and entanglement information can provide insights into the emergence of time evolution in quantized Einstein gravity. 
It introduces the notion of local time through the state of the system.
\\

\noindent
It is worth noting that the quantized Einstein gravity theory in a closed universe, described by the Wheeler-DeWitt equation, does not naturally incorporate time evolution. 
However, modular evolution, which relies on the entanglement information of a state, can provide insights into the emergence of time evolution. 
This is where the connection between the LEHM (modular parameter) and the Out-of-Time-Ordered Correlator (OTOC) duality becomes relevant. 
The duality between the LEHM and OTOC suggests that the modular parameter is related to the notion of time evolution. 
Therefore, the idea put forward is that time evolution should emerge from the modular flow, entanglement, and chaos bounds. 
\\

\noindent
The entanglement spectrum, which represents the spectrum of eigenvalues of a reduced density matrix, can capture important features of a system's dynamics and correlations. 
In certain cases, it has been observed that the entanglement spectrum can capture important features of the system's dynamics and correlation functions \cite{Schmitt:2018diw,Yan:2019fbg,Bhattacharyya:2019txx}. 
It can be related to correlation functions through the bulk reconstruction correspondence, allowing one to obtain information about correlations and dynamics in the bulk of the system. 
Analogies have been drawn between the eigenvalues of a modular Hamiltonian (which governs the entanglement structure) and the frequencies of a drum. 
Just as different frequencies of a drum produce different vibrational patterns, the eigenvalues of a modular Hamiltonian can be associated with different entanglement patterns in the system. 
Furthermore, the eigenvectors of a subsystem Hamiltonian (which can be obtained from the entanglement spectrum \cite{Denton:2019pka}) can be likened to the shape of a drum. 
\\

\noindent
In this letter, we explore a specific set of random operators and their connection to two important concepts in quantum physics: the Loschmidt echo (LE) and the SFF. 
Moreover, we extend the existing studies on bosonic systems \cite{Yan:2019fbg,Bhattacharyya:2019txx,deMelloKoch:2019rxr,Ma:2019ocx} to the realm of fermionic systems. 
The thermofield double (TFD) state and Berry phase \cite{Nogueira:2021ngh,Huang:2021qkm,Banerjee:2022jnv} serve as our chosen applications for exploring the implications of these random operators. 
To establish our results, we rely on the concept of unitary equivalence, which imposes constraints on the subsystem Hamiltonian $\hat{H}$ and the modular Hamiltonian $\hat{H}_{\mathrm{mod}}$. To delve into our analysis, we employ the eigenvalue-eigenvector identity. 
This identity allows us to use the entanglement spectrum as a tool for reconstructing the Hamiltonian of the system. 

\section{SFF and LE in a Fermionic System}
\label{sec:2} 
\noindent
This section considers a generalization of Refs. \cite{Yan:2019fbg,Bhattacharyya:2019txx} to the fermionic case. We will first lay out the setup and show the similarity between the fermionic and bosonic systems. Then we will demonstrate how the SFF and LE can be related to a specific type of correlators by integrating over the parameters in fermionic coherent states.

\subsection{Set-Up}
\noindent
Here, we briefly review the fermionic coherent state and introduce the relevant operator in the calculations. 
We consider a two-state system $(\left| 0 \right\rangle, \left| 1 \right\rangle \equiv \hat{c}^\dagger\left| 0 \right\rangle)$ with a pair of creation and annihilation operators $(\hat{c}^\dagger, \hat{c})$ satisfying the usual anti-commutation relation $\{\hat{c}, \hat{c}^{\dagger}\}=1$. 
The eigenstates correspond to the eigenvalues of the number operator $\hat{N}=\hat{c}^\dagger \hat{c}$, i.e. $\hat{N}\left| 0 \right\rangle=0$ and $\hat{N} \left| 1 \right\rangle =\left| 1 \right\rangle$. 
\\

\noindent
The fermionic coherent state with Grassmann-valued parameters $(\psi, \psi^*)$ is defined by
\bea
|\psi\rangle\equiv \hat{D}(\psi)|0\rangle
=\bigg(1-\frac{1}{2}\psi^*\psi\bigg) |0\rangle-\psi |1\rangle,
\eea 
which is generated by a displacement operator $\hat{D}\equiv \exp(\hat{c}^{\dagger}\psi-\psi^* \hat{c})$ acting on the vacuum state.
\\


\noindent
The operators (defined at time $t=0$ in the Heisenberg picture) which are central to our calculation are defined by
\bea
\hat{U}_{(\alpha,\gamma)}(0) \equiv e^{i(\alpha^*\hat{c}+\hat{c}^{\dagger}\alpha)
+(\gamma^*\hat{t}-\hat{t}^{\dagger}\gamma)}e^{-\frac{1}{2}\alpha^*\alpha}, 
\eea
where 
\bea
\hat{t}\equiv\frac{\partial}{\partial\psi^*}; \qquad  \hat{t}^{\dagger}\equiv\frac{\partial}{\partial \psi}
\eea 
and Grassmann-valued parameters ($\alpha$,$\gamma$). 
\\
\\
\noindent
With the definition of the operator, it is equivalent to separating into two parts $\hat{U}(0)= \hat{S}(0)\hat{T}(0)$ using the Baker-Campbell-Hausdorff formula, where 
\bea
\hat{S}(0)\equiv e^{i\alpha^{*}\hat{c}}e^{i\hat{c}^{\dagger}\alpha}, \  
\hat{T}(0)\equiv e^{\gamma^*\hat{t}-\hat{t}^{\dagger}\gamma}. 
\eea  
The $\hat{T}$ operator acts on the coherent state as a shift to the parameter:
\bea
\hat{T}(0)|\psi\rangle=|\psi+\gamma\rangle, \ 
\langle\psi|\hat{T}^{\dagger}(0)=\langle\psi+\gamma|. 
\eea
\noindent
This operator $\hat{U}$ is inspired by the group element of the Heisenberg group when one considers a bosonic quantum mechanical system \cite{deMelloKoch:2019rxr,Ma:2019ocx}
\bea
\hat{U}(q_1, q_2)\equiv\exp(iq_1\hat{X}(0)+iq_2\hat{P}(0)),
\eea 
where $q_1$, $q_2$ are bosonic parameters, and $\hat{X}(0)$ and $\hat{P}(0)$ are position and momentum operators, respectively. 
The operators act on the eigenstate of $\hat{X}(0)$ as 
\bea
e^{iq\hat{X}(0)}|x\rangle=e^{iqx}|x\rangle, \ 
e^{iq\hat{P}(0)}|x\rangle=|x-q\rangle. 
\eea  
Due to the similarity between the roles of $\hat{X}$, $\hat{P}$ and $\hat{S}$, $\hat{T}$, we can generalize the bosonic result to a fermionic system in a similar fashion.

\subsection{Spectral Form Factor} 
\noindent
For a system described by a Hamiltonian $\hat{H}$ at inverse temperature $\beta$, the partition function is given by:
\begin{align}
    Z(\beta) &= \Tr e^{-\beta \hat{H}} = \left\langle e^{-\beta \hat{H}} \right\rangle = \int d\psi^* d\psi \left\langle \psi \right| e^{-\beta \hat{H}} \left| \psi \right\rangle \,.
\end{align}
The spectral form factor (SFF) is the absolute value square with a complex inverse temperature $S_2(\beta, t)\equiv |Z(\beta+it)|^2$. 
The $t$ is the time of the system. 
The SFF captures information about the spectrum of the system and should be a probe to quantum chaos at a late time \cite{Dyer:2016pou}. 
We will show that the SFF at inverse temperature $\beta/2$ is equivalent to the thermal expectation value of a two-point function averaged over the coherent state and the operators $\hat{U}$ of the form:
\begin{align}
G_{AV, 2} &= \int d\psi^{*}d\psi d\alpha^{*}d\alpha d\gamma^{*}d\gamma\ 
\langle\psi|e^{-\frac{\beta}{2}\hat{H}}\hat{U}_{(\alpha,\gamma)}(t)e^{-\frac{\beta}{2}\hat{H}}\hat{U}_{(\alpha,\gamma)}^{\dagger}(0)|\psi\rangle \nn\\
&= \int_{\psi} \int_{\alpha,\gamma} 
\langle\psi|e^{-\frac{\beta}{2}\hat{H}}\hat{U}_{(\alpha,\gamma)}(t)e^{-\frac{\beta}{2}\hat{H}}\hat{U}_{(\alpha,\gamma)}^{\dagger}(0)|\psi\rangle
\,.
\end{align}
We have also used a shorthand notation
\begin{align}
\int_\psi \int_{\alpha, \gamma} \equiv \int d\psi^{*}d\psi d\alpha^{*}d\alpha d\gamma^{*}d\gamma \,.
\end{align}
Note that we have regularized the thermal expectation value by inserting the thermal factors slightly differently \cite{Liao:2018uxa}. 
Let us proceed with the calculation while suppressing the parameters of the $\hat{U}$ operators:
\begin{align}
G_{AV, 2} &= \int_\psi \int_{\alpha, \gamma} 
\langle\psi|e^{-\frac{\beta}{2}\hat{H}}\hat{U}(t)e^{-\frac{\beta}{2}\hat{H}}\hat{U}^{\dagger}(0)|\psi\rangle \nn\\
&= 
\int_\psi \int_{\alpha, \gamma} 
\langle\psi|e^{(it-\frac{\beta}{2})\hat{H}}\hat{T}(0)\hat{S}(0)e^{-(it+\frac{\beta}{2})\hat{H}}\hat{S}^{\dagger}(0)\hat{T}^{\dagger}(0)|\psi\rangle  \nn\\
&=
\int_{\substack{\psi,\psi_1 \\ \psi_2 , \psi_3} } \int_{\alpha, \gamma}\langle\psi|e^{(it-\frac{\beta}{2})\hat{H}}|\psi_1\rangle\langle\psi_1|
\hat{T}(0)e^{i\alpha^*\hat{c}}|\psi_2\rangle\langle\psi_2| 
e^{i\hat{c}^{\dagger}\alpha}e^{-(it+\frac{\beta}{2})\hat{H}}e^{-i\alpha^*\hat{c}}|\psi_3\rangle \nn \\
& \quad \times\langle\psi_3|
e^{-i\hat{c}^{\dagger}\alpha}\hat{T}^{\dagger}(0)|\psi\rangle \nn \\
&=
\int_{\substack{\psi,\psi_1 \\ \psi_2 , \psi_3} } \int_{\alpha, \gamma} 
e^{i\alpha^*(\psi_2-\psi_3)}e^{i(\psi_2^*-\psi_3^*)\alpha} 
\langle\psi|e^{(it-\frac{\beta}{2})\hat{H}}|\psi_1\rangle 
\langle\psi_1|\hat{T}(0)|\psi_2\rangle
\langle\psi_2|e^{-(it+\frac{\beta}{2})\hat{H}}|\psi_3\rangle 
\langle\psi_3|\hat{T}^{\dagger}(0)|\psi\rangle \label{eqn:sff_1}
\\
&= 
\int_{\substack{\psi,\psi_1 \\ \psi_2 , \psi_3} } \int_{\gamma} 
\delta(\psi_2-\psi_3)
\langle\psi|e^{(it-\frac{\beta}{2})\hat{H}}|\psi_1\rangle 
\langle\psi_1|\hat{T}(0)|\psi_2\rangle 
\langle\psi_2|e^{-(it+\frac{\beta}{2})\hat{H}}|\psi_3\rangle 
\langle\psi_3|\hat{T}^{\dagger}(0)|\psi\rangle \label{eqn:sff_2} \\
&=
\int_{\psi, \psi_1,\psi_2 } \int_{\gamma}
\langle\psi|e^{(it-\frac{\beta}{2})\hat{H}}|\psi_1\rangle 
\langle\psi_1|\hat{T}(0)|\psi_2\rangle 
\langle\psi_2|e^{-(it+\frac{\beta}{2})\hat{H}}|\psi_2\rangle 
\langle\psi_2|\hat{T}^{\dagger}(0)|\psi\rangle \nn\\
&=
\int_{\psi, \psi_1,\psi_2 } \int_{\gamma} \langle\psi|e^{(it-\frac{\beta}{2})\hat{H}}|\psi_1\rangle 
\langle\psi_1|\psi_2+\gamma\rangle  \langle\psi_2+\gamma|\psi\rangle
\langle\psi_2|e^{-(it+\frac{\beta}{2})\hat{H}}|\psi_2\rangle 
 \label{eqn:sff_3} \\ 
&= \int_\psi \langle\psi|e^{-(\frac{\beta}{2}-it)\hat{H}}|\psi\rangle  \int_{\psi_2} \langle\psi_2|e^{-(\frac{\beta}{2}+it)\hat{H}}|\psi_2\rangle \nn\\
&= \left|Z\left(\beta/2 + it\right)\right|^2 = S_2(\beta/2 , t) \,.
\end{align}
In the calculation above, we have inserted three complete sets of coherent states denoted by $\psi_1$, $\psi_2$, and $\psi_3$ in Eq. \eqref{eqn:sff_1}. The integration over $\alpha$ in Eq. \eqref{eqn:sff_2} generates the Grassmann delta function. 
Finally, the integrations in Eq. \eqref{eqn:sff_3} over $\gamma$ and $\psi_1$ give identity:
\begin{align}
    &\int_\gamma \int_{\psi_1}  |\psi_1\rangle \langle\psi_1|\psi_2+\gamma\rangle \langle\psi_2+\gamma|\psi\rangle \nn\\
    &= \left(\int_{\psi_1}  \left|\psi_1 \right\rangle \left\langle\psi_1 \right| \right) \left( \int_\gamma \left|\psi_2+\gamma\right\rangle \left\langle\psi_2+\gamma \right| \right) \left|\psi \right\rangle \nn\\
    &= \left|\psi\right\rangle.
\end{align}
This calculation shows that the regularized thermal expectation value of 
\bea
\left\langle\hat{U}_{\alpha,\gamma}(t) \hat{U}_{\alpha,\gamma}(0)\right\rangle_\beta
\eea
 at inverse temperature $\beta$ averaged over the states and parameters in the operator $\hat{U}$ is equivalent to SFF at inverse temperature $\beta/2$. 
We can immediately extend this result to higher point functions and the corresponding SFF with adjusted inverse temperature. 
A useful equation for the generalization is
\begin{align}
\int_{\alpha, \gamma}\ \langle\psi|\hat{U}(0) e^{i\hat{H}t}\hat{U}^{\dagger}(0)|\psi_1\rangle =\langle\psi|\psi_1\rangle\langle e^{i\hat{H}t}\rangle. 
\label{idh}
\end{align}
This equation also arose from the Haar average of the unitary operator employed in Refs. \cite{Yan:2019fbg,Bhattacharyya:2019txx}. 
With Eq. \eqref{idh}, it is not difficult to generalize the above calculation to $2k$-point correlation functions. 
The result is
\begin{align}
&\int_{\psi} \int_{ \{ \alpha_j, \gamma_j \} } \langle\psi| \hat{U}_1^0\rho_\beta^{\frac{1}{2k}} \hat{U}_2^t\rho_\beta^{\frac{1}{2k}}\hat{U}_3^0\cdots 
\hat{U}_{2k-2}^t\rho_\beta^{\frac{1}{2k}}\hat{U}_{2k-1}^0
\rho_\beta^{\frac{1}{2k}}\hat{W}_{2k}(t)\rho_\beta^{\frac{1}{2k}}|\psi\rangle
\nn\\
&=
|\langle e^{(-\frac{\beta}{2k}+it)\hat{H}}\rangle|^{2k}
, 
\end{align}
where $\rho_\beta^\frac{1}{2k} \equiv \big(\exp(-\beta \hat{H})\big)^{\frac{1}{2k}}$ is the regularized thermal factor, 
\bea
\hat{W}_{2k-1}(t)\equiv\big(\hat{U}_1(t)\hat{U}_2(t)\cdots\hat{U}_{2k-1}(t)\big)^{\dagger},
\eea
and we have used superscript to denote the time of the operator insertion and subscript to denote the parameters (of the operator), e.g. $\hat{U}^t_2 \equiv \hat{U}_{\alpha_2,\gamma_2} (t)$. 
Note that for the higher-point function, the insertion of the operators is out-of-time-ordered. 
For example, in the case of $k=2$, the relation is 
\begin{align}
 	&\int_{\psi} \int_{ \{ \alpha_j, \gamma_j \} } \langle\psi| \hat{U}_1(0)\rho_\beta^{\frac{1}{4}} \hat{U}_2(t)\rho_\beta^{\frac{1}{4}}\hat{U}_3(0) \rho_\beta^{\frac{1}{4}} \hat{W}_{4}(t)\rho_\beta^{\frac{1}{4}}|\psi\rangle = 	|\langle e^{(-\frac{\beta}{4}+it)\hat{H}}\rangle|^{4} \, , 
\end{align}
where $\hat{W}_4(t) \equiv \big(\hat{U}_1(t)\hat{U}_2(t)\hat{U}_{3}(t)\big)^{\dagger}$. 
Hence the above relation can be interpreted as a connection between a specific kind of out-of-time-order correlator (OTOC) at inverse temperature $\beta$ and the spectral form factor at a reduced inverse temperature.
 
 \subsection{Loschmidt Echo} 
\noindent
In the previous section, we studied the relationship between the OTOC and SFF. 
The OTOC captures the growth of the deviation of a small perturbation on the initial condition. 
The SFF captures the characteristic distribution of the spectrum of a chaotic system. 
It turns out that these two quantities are also dual to another measurement (for diagnosing chaos) called Loschmidt Echo, $M(t)$,
\begin{align}
	M(t) &\equiv \left|\left\langle \psi_0 \right| e^{i \hat{H}_2 t} e^{-i \hat{H}_1 t} \left| \psi_0 \right\rangle \right|^2 \,.
\end{align}
From the definition, it is similar to measuring the deviation of an initial state $\left|\psi_0\right\rangle$ after a time evolution forward in time with a Hamiltonian $\hat{H}_1$ and then reversing time evolution with another Hamiltonian $\hat{H}_2$. 
When the two Hamiltonians differ only by a small perturbation 
\bea
\hat{H}_2= \hat{H}_1 + \delta \hat{H} \ M(t) \sim \left|\left\langle \psi_0 \right| \exp(i \delta \hat{H} t) \left| \psi_0 \right\rangle \right|^2 ,
\eea
it measures the deviation of growth. 
The LE should be related to the OTOC as they capture similar information. 
\\

\noindent
Next, we will demonstrate the connection between the OTOC correlators and LE. 
The calculation for the fermionic system is in line with the bosonic case \cite{Yan:2019fbg,Bhattacharyya:2019txx}. 
Let us consider a lattice system with $k$ sites equipped with a set of non-trivial operators $\{\hat{U}_j \}$, $j=1,\cdots,k$ localized on each site labeled by the subscript (with a tensor product of the identity operator at other sites). 
With this setup, let us consider the following $2k$-point correlator
\begin{align}
&G_{AV, 2k}
\nn\\
&=
\int _\psi \int_{U_1 \cdots U_k}\
\langle\psi|\hat{U}_1^{\dagger}(t_1)\hat{U}_2^{\dagger}(t_2)\cdots \hat{U}_{k-1}^{\dagger}(t_{k-1})\hat{U}_k^{\dagger}(0)
\hat{U}_{k-1}(t_{k-1})\cdots \hat{U}_1(t_1)\hat{U}_k(0)|\psi\rangle \,,
\end{align}
where we insert the operator $\hat{U}_j(t_j)$ localized on the $j$-th site at time $t_j$. 
Here we choose $t_{k-1} > t_{k-2} > \cdots > t_1 > t_k=0$, which reproduces the same time ordering in the conventional OTOC. 
Note that we have also used a shorthand notation $\int_{U}\equiv\int d\hat{U}$ to denote the integration over the operator parameters used in the previous section. 
We will use a normalized integration measure such that
\bea
\int d\hat{U}_j(0)=1. 
\eea
When $k=1$, the correlator is trivial and gives unity. 
The non-trivial result begins from $k=2$, which takes the form of the standard four-point OTOC:
\bea
G_{AV, 4}
&=&
\int_\psi \int_{U_1,U_2} \, 
\langle\psi| \hat{U}_1^{\dagger}(t_1) \hat{U}_2^{\dagger}(0)
\hat{U}_1(t_1)\hat{U}_2(0)|\psi\rangle
\nn\\
&=&
\int_\psi \int_{U_1,U_2} \, 
\langle\psi|e^{i\hat{H}t_1}\hat{U}_1^{\dagger}(0)e^{-i\hat{H}t_1}\hat{U}_2^{\dagger}(0)
e^{i\hat{H}t_1}\hat{U}_1(0)e^{-i\hat{H}t_1}\hat{U}_2(0)|\psi\rangle
\nn\\
&=&
\int_{\psi, \hat{U}_2}\ 
\langle\psi|e^{i\hat{H}t_1}
 \left[\int_{U_1} \, \hat{U}_1^\dagger(0) e^{-i\hat{H}t_1}\hat{U}_2^{\dagger}(0)
e^{i\hat{H}t_1} \hat{U}_1(0)\right]
e^{-i\hat{H}t_1}\hat{U}_2(0)|\psi\rangle
\nn\\
&=&
\int_{\psi, \hat{U}_2}\ 
\langle\psi|
\Tr_1 \left[e^{-i\hat{H}t_1}\hat{U}_2^{\dagger}(0)
e^{i\hat{H}t_1} \right]
e^{-i\hat{H}t_1}\hat{U}_2(0) e^{i\hat{H}t_1}|\psi\rangle,
\eea
where we replaced the average over the operator $\hat{U}_1$ by a trace over site 1 \cite{Yan:2019fbg,Bhattacharyya:2019txx}.
To proceed, let us assume the total system Hamiltonian takes the form of 
\bea
\hat{H}= \hat{H}_1 \otimes \hat{I}_2 - \hat{I}_1 \otimes \hat{H}_2 + \sum_\alpha \hat{H}_\alpha'\otimes \hat{H}_\alpha''.
\eea 
The interaction terms are assumed to be random such that a partial trace operation leads to a sum of operators (denoted as $\hat{P}_1$ and $\hat{P}^{\prime}_1$ below). 
For further detail on the properties of the random interacting Hamiltonian, please refer to Ref. \cite{Yan:2019fbg}. 
With this assumption, the evaluation of trace over site one is: 
\begin{align}
\Tr_1 \left[e^{-i\hat{H}t_1}\hat{U}_2^{\dagger}(0)
e^{i\hat{H}t_1} \right] &= \int_{\psi_1}\ \langle\psi_1| e^{-i\hat{H}t_1}\hat{U}^{\dagger}_2(0)e^{i\hat{H}t_1}|\psi_1\rangle \nn\\
&=\frac{1}{N_1}\sum_{\hat{P}_1}e^{-i(\hat{H}_2+\hat{P}_1)t_1}\hat{U}_2^{\dagger}(0)e^{i(\hat{H}_2+\hat{P}_1)t_1} \,,  
\end{align}
where $\psi_1$ is the complete set of states in site 1, and the constant $N_1$ is the number of operators $\hat{P}_1$. 
Hence we obtain that:
\bea
&&
G_{AV, 4}
\nn\\
&=&\frac{1}{N_1}\int_{\psi_1,\psi_2} \int_{\hat{U}_2}\ \left\langle\psi_2\right|
\sum_{\hat{P}_1, \hat{P}'_1}e^{-i(\hat{H}_2+\hat{P}_1)t_1}\hat{U}_2^{\dagger}(0)e^{i(\hat{H}_2+\hat{P}_1)t_1}
\nn\\
&&\times
 \left[\left\langle\psi_1 \right|
e^{-i\hat{H}t_1}\hat{U}_2(0)e^{i\hat{H}t_1} \left| \psi_1 \right\rangle \right] \left|\psi_2\right\rangle
\nn\\
&=&\frac{1}{N_1^2}\int_{\psi_2} \int_{\hat{U}_2}\, \left\langle\psi_2\right|
\sum_{\hat{P}_1, \hat{P}'_1}e^{-i(\hat{H}_2+\hat{P}_1)t_1}\hat{U}_2^{\dagger}(0)e^{i(\hat{H}_2+\hat{P}_1)t_1}
e^{-i(\hat{H}_2+\hat{P}'_1)t_1}\hat{U}_2(0)
\nn\\
&&\times
e^{i(\hat{H}_2+\hat{P}'_1)t_1}\left|\psi_2\right\rangle
\nn\\
&=&\frac{1}{N_1^2}\int_{\psi_2} \left\langle\psi_2\right|
\sum_{\hat{P}_1, \hat{P}'_1}e^{-i(\hat{H}_2+\hat{P}_1)t_1} \left[\int_{\hat{U}_2}\,\hat{U}_2^{\dagger}(0)e^{i(\hat{H}_2+\hat{P}_1)t_1}
e^{-i(\hat{H}_2+\hat{P}'_1)t_1}\hat{U}_2(0)\right] 
\nn\\
&&\times 
e^{i(\hat{H}_2+\hat{P}'_1)t_1}\left|\psi_2\right\rangle
\nn\\
&=&
\frac{1}{N_1^2}\sum_{\hat{P}_1, \hat{P}'_1} \, \left|\int_{\psi_2}\left\langle\psi_2\right|e^{i(\hat{H}_2+\hat{P}_1)t_1}e^{-i(\hat{H}_2+\hat{P}'_1)t_1}\left|\psi_2\right\rangle\right|^2.  
\eea
We have split the whole system state into the tensor product states in site 1 and site 2, $\left|\psi \right\rangle = \left|\psi_1\right\rangle \otimes \left|\psi_2 \right\rangle$, from the first to the second line. 
The averaged four-point correlation function is equivalent to LE with a summation of different time evolution operators. 
When applying a late-time limit, only the greatest exponent will survive. 
Therefore, we can apply this result to modular chaos \cite{DeBoer:2019kdj} after applying the unitary equivalence \cite{Jafferis:2020ora}. 
We can also use the approximation of Ref. \cite{Bhattacharyya:2019txx} to generalize this result to high-point correlators. 
The calculation is similar to SFF. 

\subsection{Field Theory Generalization}
\noindent 
It is easy to generalize the above calculations to multi-fermion systems. 
The corresponding multi-fermion coherent states and the operators $\hat{U}$ are given by:
\bea
|\vec{\psi}\rangle
&\equiv&|\psi_1, \psi_2, \cdots, \psi_n\rangle
=\hat{D}(\psi_1, \psi_2, \cdots, \psi_n)|\vec{0}\rangle
\nn\\
&=&\prod_{j=1}^n\bigg(1-\frac{1}{2}\psi^*_j\psi_j-\psi_j\hat{c}^{\dagger}_j\bigg)\bigg|\vec{0}\bigg\rangle; 
\eea
\bea
&&
\hat{U}(0)=\hat{S}(0)\hat{T}(0), \ 
 \hat{S}(0)\equiv\prod_{j=1}^n e^{i\alpha_j^{*}\hat{c}_j}e^{i\hat{c}_j^{\dagger}\alpha_j}, \ 
\hat{T}(0)\equiv\prod_{j=1}^n e^{\gamma_j^*\hat{t}_j-\hat{t}_j^{\dagger}\gamma_j}, 
\eea
where $n$ is the number of fermion fields. The calculation follows similarly as in the single fermion case.

\section{Connection between Time Evolution and Modular Flow through the TFD State}
\label{sec:3}
\noindent
We will discuss the relevance of the above quantum chaos diagnostic quantities to the emergence of time. 
In particular, we will motivate the connection between the time evolution generated by the subsystem Hamiltonian and the concept of modular flow generated by the modular Hamiltonian. 
We will follow the Ref. \cite{Jafferis:2020ora} and demonstrate this connection through the thermofield double state. 
The TFD state that we consider is:
\begin{align}
\left| \Psi \right\rangle &\sim \sum_m e^{-\frac{\beta}{2}E_m}\hat{U}|m\rangle\otimes|m\rangle \,, \nn\\ 
\rho_{TFD} &= \left| \Psi \right\rangle \left\langle \Psi \right| \,,
\end{align}
where it is up to a normalization constant. 
The $E_m$ is the eigenvalue of $\hat{H}$ for the state $|m\rangle$.
The $\hat{U}$ is a unitary operator which only acts on the first site. 
We can use the state for bosonic and fermionic fields corresponding to a different range of $m$. 
The Hamiltonian is $\hat{H}=\hat{H}_0\otimes\hat{I}-\hat{I}\otimes\hat{H}_0$, where $\hat{I}$ is an identity operator. 
The two systems do not interact with each other. With this definition of the TFD state, one can show a relation between the subsystem Hamiltonian, $H_0$, with the modular Hamiltonian defined through partial trace over the second site of the TFD density matrix $\hat{H}_{\text{mod}} \equiv - \ln \Tr_2 \rho_{TFD}$. 
For this case, we have the unitary equivalence in general for the TFD state, $\hat{H}_{\mathrm{mod}}=\beta\hat{U}\hat{H}_0\hat{U}^\dagger$ \cite{Jafferis:2020ora}. 
In a quantum system, the Hamiltonian generates a time evolution of the operators (in the Heisenberg picture), i.e. $\hat{{\cal O}}(t) = \exp(i\hat{H}_0t) \hat{{\cal O}}(0) \exp(-i\hat{H}_0t)$. 
On the other hand, the modular Hamiltonian allows one to define a modular flow of an operator $\hat{{\cal O}}(s) = \exp(i \hat{H}_{\text{mod}} s) \hat{{\cal O}}(0) \exp(-i \hat{H}_{\text{mod}} s)$. 
Using the unitary equivalent relation between the Hamiltonian and modular Hamiltonian, one can relate the modular flow to the time evolution of the operators by identifying $s = t/\beta$. 
This identification coincides with the "Thermal Time Hypothesis" in Ref. \cite{Connes:1994hv} where the time evolution is the modular flow (scaled by the temperature). 
We will apply this identification to the above quantum chaos diagnostics and argue how time can emerge from quantum entanglement. 
\\

\noindent 
When considering $\beta=0$, the TFD state is equivalent to summing over a complete basis (of states), either in the form of eigenstates of the number operator or coherent states, which we have considered in the previous section. 
Therefore, we can apply the result of the previous section to the TFD state.  
The four-point function for the TFD state is
\begin{align}
   G_{4}&=\left\langle \Psi \right| \hat{{\cal O}}_1(t_1) \cdots \hat{{\cal O}}_{4}(t_{4})  \left| \Psi \right\rangle. 
\end{align}
One can connect this correlation function to the regularized thermal expectation value considered in the previous section first by shifting the time parameters as 
\bea
t_j \rightarrow t_j - i \frac{\beta}{4}(2-j)\ \forall j= 1,\cdots ,4
\eea 
and then fix the operators and their insertion time accordingly. 
We choose 
\bea
\hat{{\cal O}}_1^\dagger(t_1)=\hat{{\cal O}}_3(t_3)=\hat{U}_1(t); \ 
\hat{{\cal O}}_2^\dagger(t_2)=\hat{{\cal O}}_4(t_4)=\hat{U}_2(0).
\eea 
Let us recall the corresponding result for the LE in the previous sections, LE is given by
\begin{align}
G_{AV, 4} = \frac{1}{N_1^2}\sum_{\hat{P}_1, \hat{P}'_1} \, \left|\int_{\psi_2}\left\langle\psi_2\right|e^{i(\hat{H}_2+\hat{P}_1)t_1}e^{-i(\hat{H}_2+\hat{P}'_1)t_1}\left|\psi_2\right\rangle\right|^2 \,,
\label{eqn:OTOC_LE}
\end{align} 
expressed in terms of the subsystem Hamiltonian and the physical time parameter.
We then introduce a modular parameter $s\equiv t/\beta$ and replace the subsystem Hamiltonian with the modular Hamiltonian in the formula of LE using unitary equivalence. 
With this replacement, one can interpret the LE as generated by the modular Hamiltonian and only depend on the modular flow parameter. 
This form of the LE with modular Hamiltonian coincides with the definition of modular chaos introduced in Ref. \cite{DeBoer:2019kdj}. 
They showed that there exists an upper bound for modular chaos similar to the case of chaos,
\begin{align}
	\lim_{s\rightarrow\infty}\frac{d}{ds}\ln\bigg(\big|e^{-i\hat{H}_{\mathrm{mod}}s} e^{i(\hat{H}_{\mathrm{mod}}+\delta \hat{H}_{\mathrm{mod}}) s}|\bigg) \leq 2\pi \,.
\end{align}
This result allows one to define a modular Lyapunov exponent $\lambda_{\mathrm{mod}} \leq 2\pi$. 
On the other hand, the L.H.S. of Eq. \eqref{eqn:OTOC_LE} is the OTOC with the total Hamiltonian encoding information about chaos and is bounded by 
\begin{align}
\lim_{t\rightarrow\infty}\frac{d}{dt}\ln(|\text{OTOC}|) \leq \frac{2\pi}{\beta}
\end{align}
with the Lyapunov exponent $\lambda \leq 2\pi/\beta$. 
To make our statement more precise, let us consider that chaos and modular chaos bounds are saturated, i.e. $\lambda=2\pi/\beta$ and $\lambda_{\mathrm{mod}}=2\pi$ simultaneously. 
The saturation of two bounds possibly requires one to consider holographic CFTs (or large $N$ theories) \cite{Maldacena:2015waa,DeBoer:2019kdj}. 
In our case, we did not assume the form of Hamiltonian when connecting the OTOC to the LE. 
Hence our argument can be equally applied to holographic CFTs with a small perturbation parameter (or by the large $N$ techniques) \cite{deMelloKoch:2019rxr,Ma:2019ocx}. 
One example is a spherical entangling surface in CFT, where the exponent of LE saturates the modular chaos bound \cite{Huang:2021qkm}. 
When combining this result with the relation between modular Hamiltonian and Hamiltonian through unitary equivalence, one can show ``Modular Chaos Bound = Chaos Bound''. 
We summarize the result as
\begin{align}
	\text{OTOC} &\xleftrightarrow{\text{Eq. \eqref{eqn:OTOC_LE}}} \text{LE} \xleftrightarrow{\text{UE,TTH}} \text{LEMH} \nn \\
	\lambda = \frac{2\pi}{\beta} &\xleftrightarrow{\text{\color{white}11111111111111111111\,\,1}} \lambda_{\mathrm{mod}} = 2\pi  \nn \\
	\text{time flow }``t" &\xleftrightarrow{\text{\color{white}11111111111111111111\,\,1}} \text{modular flow }``s" \,,
\end{align}
where UE and TTH denote unitary equivalent and "Thermal Time Hypothesis", respectively. 
This result also implies the emergence of a time parameter in the dual system. 
The saturation of chaos bound implies the existence of the dual AdS Einstein gravity theory \cite{Maldacena:2015waa}. 
In ``Thermal Time Hypothesis'', the time is state dependent \cite{Connes:1994hv}. 
The unitary equivalence realizes the emergence of time through Quantum Entanglement.
We demonstrate how time possibly emerges from the modular parameter in Einstein gravity theory.
Hence we expect that the unitary equivalence helps listen to the shape of bulk geometry. 
\\

\noindent
Our study is not only applicable in AdS/CFT correspondence but also to the study of Quantum Information. 
The study has applicability to Quantum Entanglement and Quantum Chaos. 
The first application is the random matrix theory.  
The dynamics of SFF helps distinguish the non-integrability and integrability from the spectrum \cite{Dyer:2016pou}. 
The second application is the classification of quantum states. 
In Ref. \cite{Nogueira:2021ngh}, they found that entanglement entropy is invariant for the TFD state for arbitrary $\hat{U}$. 
The difference between states with difference $\hat{U}$, encoded in the topological data such as the Berry phase. 
It is equivalent to classifying a state from modular Berry \cite{Huang:2021qkm} and Berry phases. 
Our study also helps clarify the relationship between the modular Berry and the Berry phases. 
Let us consider the Quantum Modular Geometric Tensor given by \cite{Huang:2021qkm}
\bea
g_{jk}^{(n)}=\sum_{m\neq n}\frac{\langle n|\partial_j\hat{H}_{\mathrm{mod}}|m\rangle \left\langle m \right| \partial_k\hat{H}_{\mathrm{mod}}|n\rangle}
{\left(E_{\mathrm{mod}}^{(n)}-E_{\mathrm{mod}}^{(m)}\right)^2},  
\eea  
where $E_{\mathrm{mod}}^{(n)}$ is the $n$-th eigenvalue of a modular Hamiltonian. 
The symmetric part gives a metric, and the anti-symmetric component provides a Modular Berry Curvature. 
Using unitary equivalence, one can obtain the Quantum Geometric Tensor by replacing the modular Hamiltonian with a Hamiltonian. 
Now the anti-symmetric component becomes the Berry Curvature. 
Therefore, unitary equivalence implies `` Modular Berry Curvature = Berry Curvature'' through the Quantum Modular Geometric Tensor. 
We can obtain phases by the integration of the Berry curvatures. 
Therefore, the modular Berry curvature and Berry curvature lead to the same phase, and hence "Modular Berry phase = Berry phase".
In Ref. \cite{Nogueira:2021ngh}, the authors only used $\hat{U}$ to transform the state but kept the Hamiltonian fixed. 
Therefore, a different choice of the unitary operator $\hat{U}$ leads to different Berry phases. 
Hence one can apply our study to classify quantum states. 

\subsection{Entanglement Spectrum and Geometry}
\noindent
Different shapes of geometry are necessary to have a distinct entanglement structure. 
Otherwise, quantum entanglement is not sufficient to determine the geometry. 
\\

\noindent
We can use a formula called \textit{eigenvector-eigenvalue identity} that connects the eigenvectors to the eigenvalues of an $n\times n$ Hermitian matrix $A$ and its minor $M_{i_2}(A)$ \cite{Denton:2019pka}
\begin{align}
	|\tilde{v}_{i_1,i_2}|^2\prod_{k=1,k\neq i_1}^n (\lambda_{i_1}(A)) - \lambda_k(A)) &= \prod_{k=1}^{n-1} (\lambda_{i_1}(A) - \lambda_k(M_{i_2})) \,.
\end{align}
Here $\tilde{v}_{i_1, i_2}$ is the $i_2$-th component of the normalized eigenvector associated with the $i_1$-th eigenvalue $\lambda_i(A)$ of $A$. The minor $M_{i_1}(A)$ is a matrix constructed by removing the ${i_1}$-th row and column of $A$ and its $i_2$-th eigenvalue is labeled by $\lambda_{i_2}(M_{i_1}(A))$. 
We adopt the convention for the eigenvalues 
\bea
\lambda_1(A)\le\cdots\le \lambda_n(A).
\eea 
When $n=2$, the element of the minor is also the eigenvalue. 
Therefore, it is easier to check the eigenvector-eigenvalue identity. 
The same set of eigenvalues (of two matrices) implies similar eigenvectors (up to a unitary transformation). 
The eigenvalues and eigenvectors are one-to-one correspondence (up to a unitary transformation). 
Taking a two-qubit system as a simple example. 
The minor of a modular Hamiltonian is the probability of the spin up or down. 
Substituting the minor to entanglement entropy shows partial entanglement (up or down). 
Therefore, we expect that the minor is one necessary ingredient of the entanglement spectrum (for constraining the subsystem Hamiltonian).  
\\

\noindent
For a more general case, one can reconstruct the eigenvectors of a modular Hamiltonian from the spectrum and its minor. 
Hence the entanglement spectrum from the modular Hamiltonian provides a unique constraint on the Hamiltonian.

\section{Outlook}
\label{sec:4}
\noindent
We worked on some interesting topics related to unitary equivalence, quantum chaos diagnostics, the AdS/CFT correspondence, and the connection between entanglement and spacetime geometry. These are active areas of research with significant implications in theoretical physics. 
The unitary equivalence \cite{Jafferis:2020ora} appears to be a key concept in our work. 
We have extended the studies on the dual of the SFF \cite{deMelloKoch:2019rxr,Ma:2019ocx} and the LE \cite{Yan:2019fbg,Bhattacharyya:2019txx} to include fermionic systems. 
By generalizing these studies to fermions, we are expanding the applicability of these concepts and potentially gaining new insights into the emergence of time and the Thermal Time Hypothesis. 
The issue of unbounded operators in field theories by noting that they appear in the exponent of exponential functions in our study. 
This allows for the generalization of our results to field theory and supports the application of the unitary equivalence to arbitrary systems. 
Furthermore, the relationship between the "Modular Chaos Bound" and the "Chaos Bound", suggests that the unitary equivalence provides additional constraints. 
This finding could have significant implications for understanding chaos and its bounds in various physical systems.
\\

\noindent
Moving on to the connection between entanglement and spacetime geometry, we explore the entanglement structure and its relationship to the underlying geometry. 
The entanglement spectrum, consisting of the eigenvalues of the reduced density matrix, has been proposed as a valuable source of information about the system's geometry. 
The reconstruction of the modular Hamiltonian from the entanglement spectrum using the eigenvector-eigenvalue identity allows for the inference of the entanglement structure and, in some cases, even the recovery of the full geometry.
  
\section*{Acknowledgments}
\noindent
We thank Xing Huang for his helpful discussion. 
CTM thanks Nan-Peng Ma for his encouragement. 
PHCL acknowledges the support from JSPS KAKENHI (Grant No. 20H01902), MEXT KAKENHI (Grant No. 21H05462), and the National Center for Theoretical Sciences during this work.
CTM acknowledges the Nuclear Physics Quantum Horizons program through the Early Career Award (Grant No. DE-SC0021892); 
YST Program of the APCTP. 
We thank the University of Warsaw for hosting ``String Math 2022'', where we presented this work. 
Discussions during the workshops, ``String theory, gravity and cosmology (SGC 2022)'' and ``String Math 2022'', were helpful for this work. 


\end{document}